*Original Article*

# The Development of a Microcontroller based Smoked Fish Machine

Aldrin Joar R. Taduran

*Department of Electrical and Electronic Engineering, College of Engineering, Tarlac State University, Tarlac, Philippines.*

*Corresponding Author : ajrtaduran@tsu.edu.ph*

***Abstract*** *- The development of a microcontroller-based smoked fish machine aims to combine and automate the boiling, smoking, and drying. The machine consists of an Arduino microcontroller, heater, spark gap igniter, stepper motor with motor driver, temperature sensor, exhaust fan, DC converter, relay module, power supply, and controller box. The main mechanism of the smoked fish machine is a stepper motor that automates the process. The machine kept the temperature at 75≤90°C to cook the fish and 60≤70°C to smoke the fish. PID is used to optimize the performance of controlling the machine's temperature. The machine used galvanized steel and food-grade material to support the heating of the fish, which helps minimize contamination and burns. The typical time of operation of the entire process using the proposed machine is up to 60 minutes.*

***Keywords*** *- Machine, Microcontroller, PID, Process, Smoked fish.*

## 1. Introduction

Fish have been a major source of food for people [1]. The global average consumption of fish and other seafood per person reached a record high of 20.5 kilograms in 2019. To ensure ample supply, techniques for breeding cultured fresh and saltwater fish in ponds [2]. However, the consumption of numerous fish species encouraged modern processing and preservation techniques [3]. These techniques help the customer get the fish in good, usable condition and will prevent the early spoilage of fish [4]. Traditional preservation methods include salting, drying, heating, and smoking the fish. Smoking fish is one of the major economic activities in the Philippines. However, due to the lack of technologies available for processing smoked fish, most people have stuck with the traditional method [5]. The traditional method requires a lot of work, takes a long time, and cannot precisely regulate the smoking conditions [6]. Additionally, traditional methods commonly include using a chimney, which is considered environmentally unfriendly due to its smoke output. To achieve quality standards, one possibility is to increase the mechanization of smokehouses and dryers [7]. Several factors must be considered when creating such a machine. The smoking temperature for "hot smoking" should range from 70 to 90 °C to allow for complete protein coagulation [8]. The quality of the salt, smoke, and heat used in this procedure needs to be tuned to produce better results [9]. Meanwhile, "cold smoking" requires 20–30°C. Additionally, smoking the fish necessitates carefully handling the protein source [10], as mishandling would reduce quality [11]. Therefore, understanding the mechanisms involved in the smoking process is crucial for maximizing the effects of smoke on fish products [12, 13]. The goal is to increase productivity, lessen reliance on manpower, and increase quality [14].

Several machines have been developed for the purpose of smoking the fish. One study integrated a blower into the kiln of a developed smoking machine and that of a traditional machine [15]. Another machine, developed by Nurmianto et al., was a mobile, portable and ergonomic fish smoke machine installed on a motorcycle. It was developed to reduce smoke to a level that is safe, comfortable, and healthful for users while also being effective for processed seafood products [16]. The said device was intended to be used for storage during transportation.

Meanwhile, Sudrajad et al. designed a milkfish preserving machine using the liquid smoke method [17]. Another device developed by Jayanti et al. included a "cold smoking" equipment with closed circulation and equipped with a temperature and concentration monitoring system based on Arduino Uno. For this machine, the fish hang vertically [18]. Odutan et al., on the other hand, optimized the loading capacity of the smoking kiln in drying catfish Clarias gariepinus [19]. Meanwhile, a multi-purpose portable smokehouse device was developed that grouped significant fish smoking methods into one process. However, the device is manually operated [20]. In line with the studies mentioned above, significant findings became a guide for the researcher for a much better design of a fish smoking machine. For one, major parts of the smoking process were integrated into the machine, namely boiling, smoking, and handling of the fish. The use of the liquid smoke method and vertically hanging the fish will ensure efficient circulation inside the chamber. Most of the literature mentioned was limited to monitoring the temperature. The researcher included a microcontroller to control the temperature for both the boiling, smoking, and drying processes. The machine is also equipped with lifting mechanisms to lessen the manual handling of fish. The machine is based on "hot smoking" temperatures and is tested for scadfish, milkfish, and tilapia.

## 2. Design Consideration

The design considerations of the model of the microcontroller-based smoked fish machine, mechanical considerations, and machine description are included in this section. The details of the machine's structure will be discussed in the design model section. The mechanism of the smoked fish machine will be discussed in the mechanical consideration section of the design-driven pulley system. The machine details the component's placement, which is significant to the process of boiling, smoking, and drying the smoked fish.

### *2.1. Design Model*

The 3D design model of a microcontroller-based automatic smoke fish machine is designed using sketchup software. This design considers the chamber's size, ventilations, and fish-rack placement, shown in Figure 1. The machine structure is divided into three primary chambers: the boiler, the dryer, and the smoker.

The machine consists of the following main components: the exhaust fan for the dryer, heat rod, fish tray, stainless bowl for the boiler, smoke path, sawdust wood platform, and igniter for the smoker. The stepper motor is the main mechanism in processing the smoked fish. The frame and structure used for this machine are galvanized sheets, iron bars, and stainless bars. This allows it to withstand more than 100 °C. Insulation is also used for the protection of electrical and electronic devices.

### *2.2. Mechanical Consideration*

The mechanism of the machine is composed of the belt, driven pulley, fish tray, and NEMA 23 stepper motor, as shown in Figure 2. This design supports the mechanism of carrying and lifting the 5kg fish. A single fixed pulley is used to carry the maximum weight of 5 kg of fish. The following equation is used to compute the lifting required for the single fixed pulley. The weight of the load (W) is computed using Equation 1.

$$\boldsymbol{W = mg} \tag{1}$$

Where m = mass and g = gravitational acceleration (9.81 m/s$^2$). The effort force for a single pulley is computed using Equation 2.

$$S = \frac{W}{P} \tag{2}$$

Where S = effort force and P = number of pulleys. This computation was used to design the pulley system of the smoked fish machine.

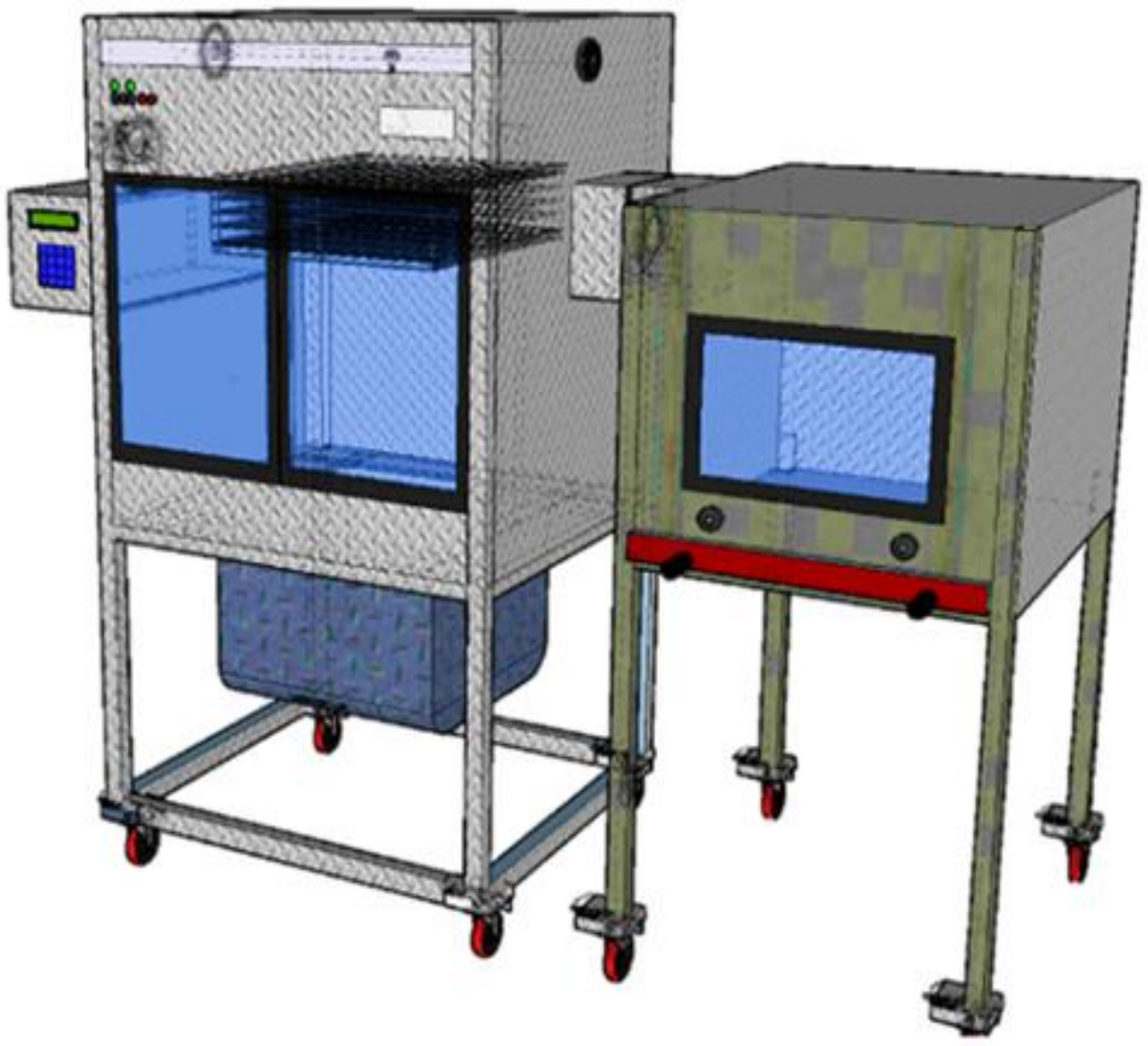

**Fig. 1 The 3D model design of the smoked fish machine**

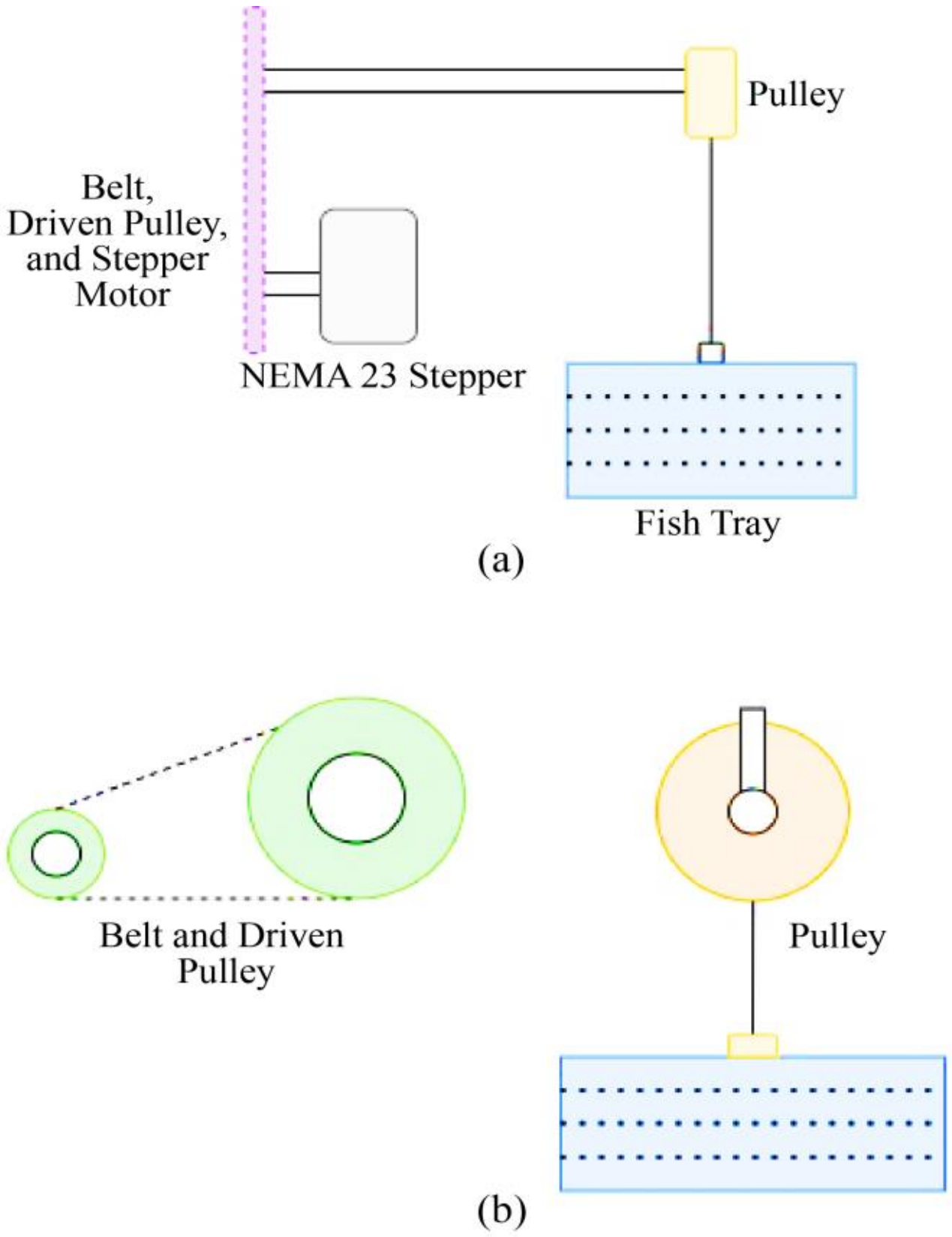

**Fig. 2 (a) Representation of mechanical mechanism of the smoked fish machine (b) Orientation of pulley and driven pulley**

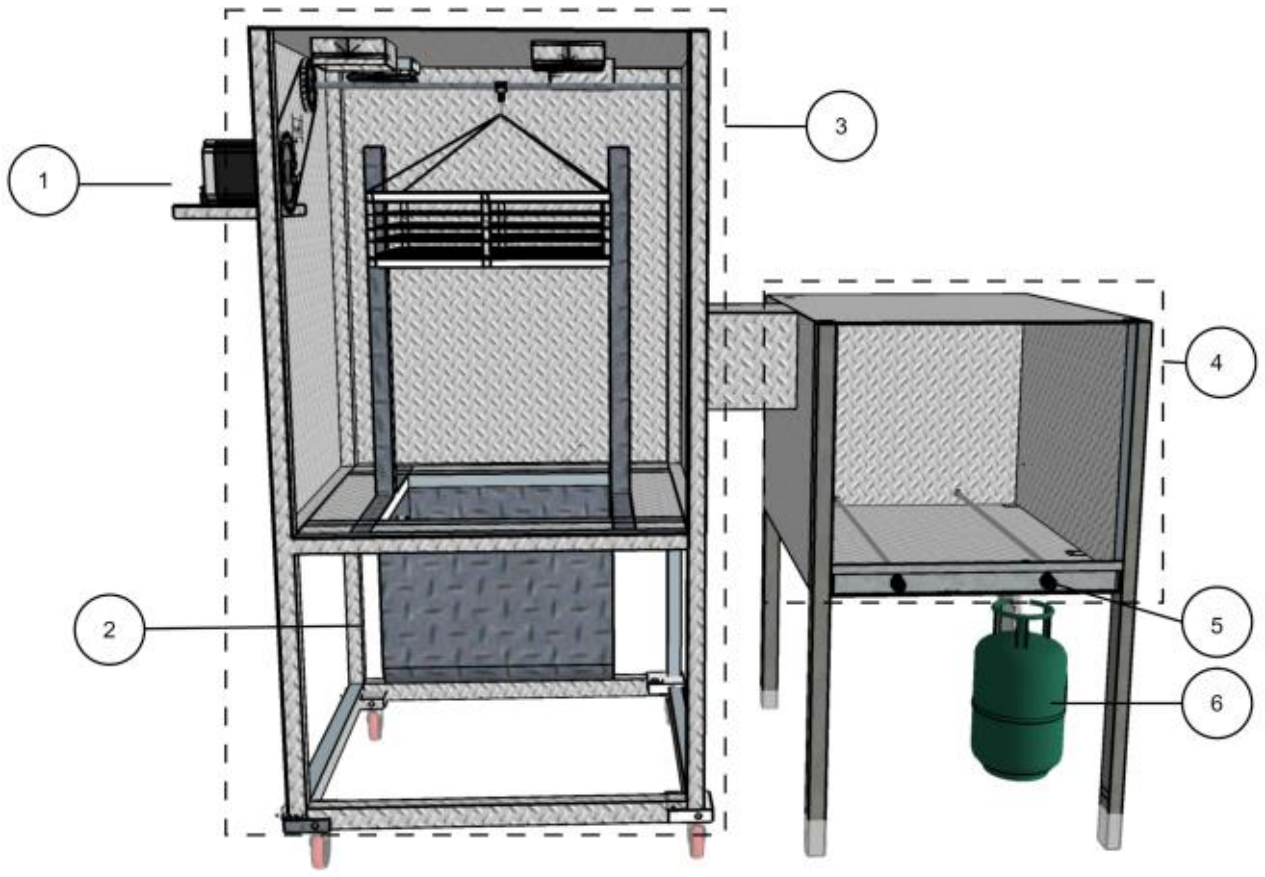

**Fig. 3 Cross-sectional view of major parts**

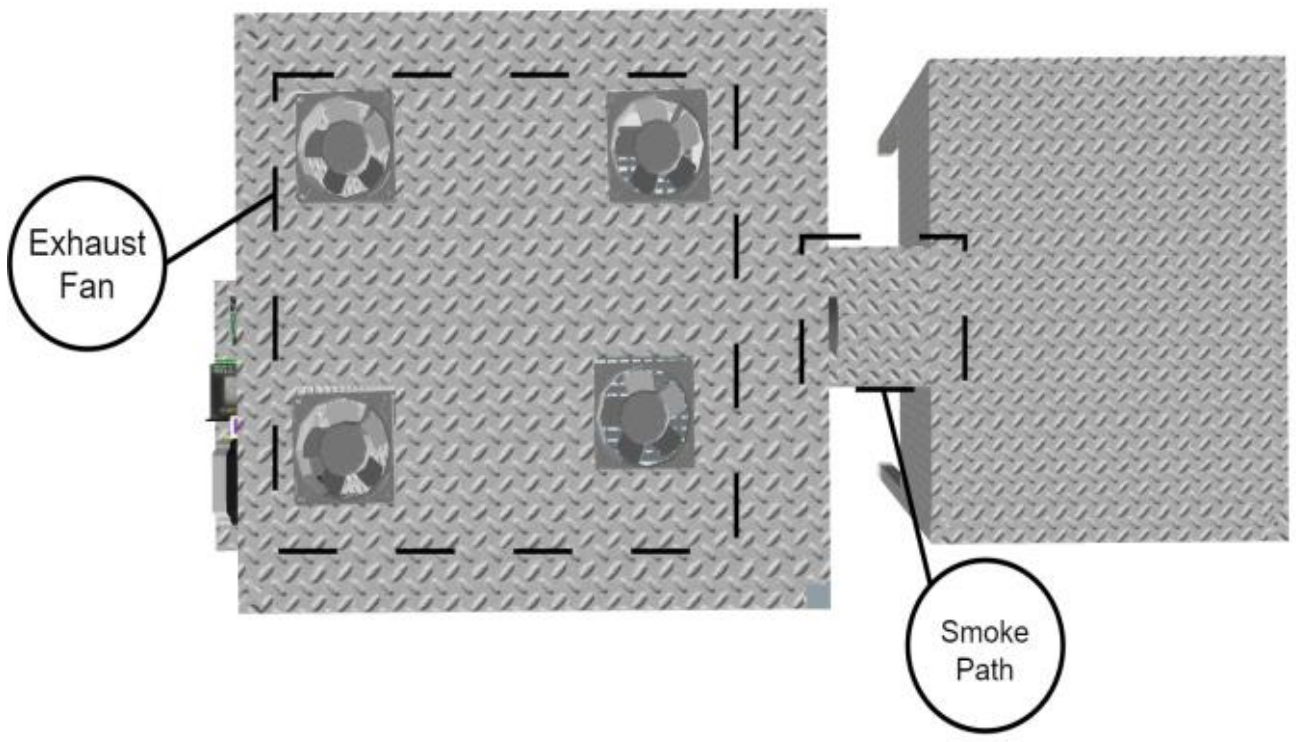

**Fig. 4 Top view of the machine**

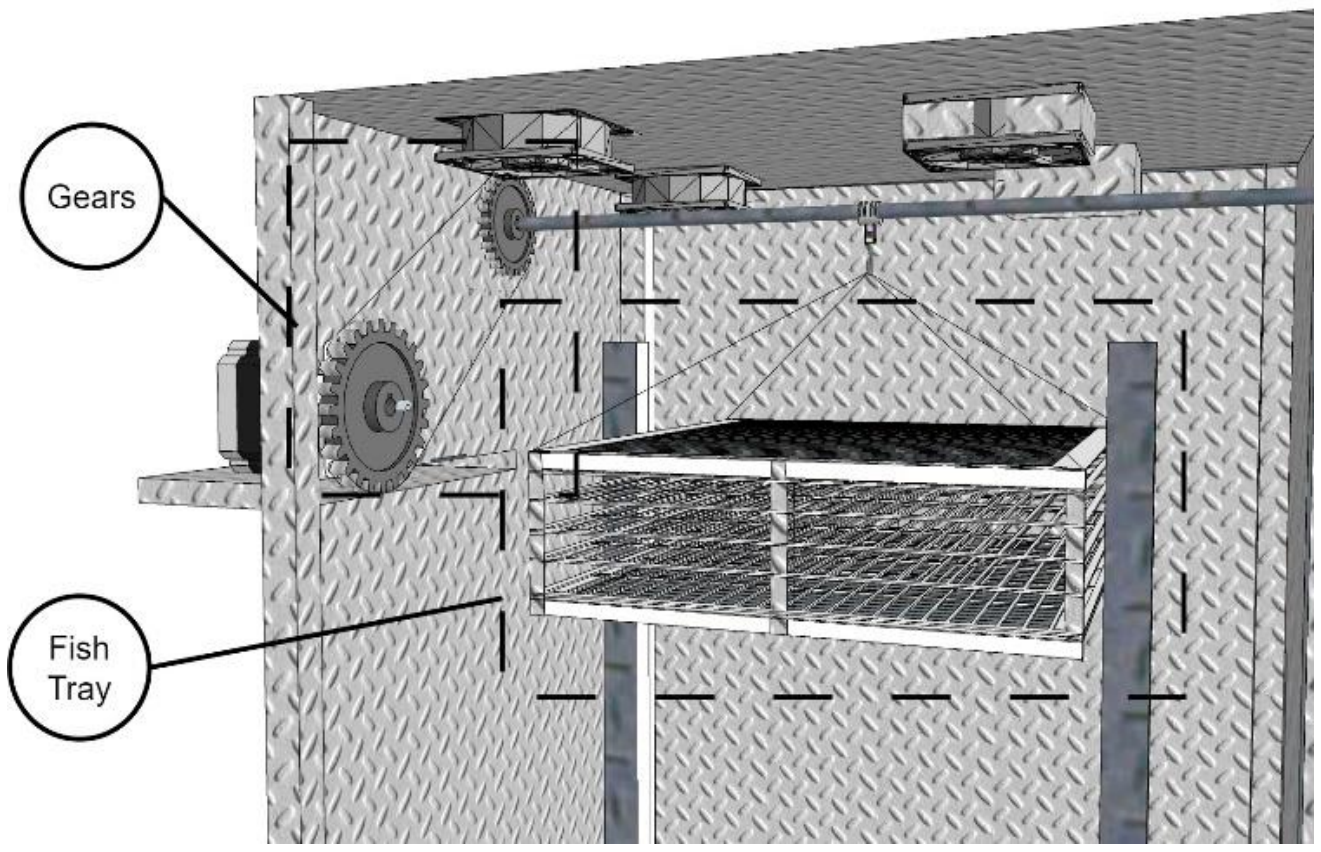

**Fig. 5 Gears and a fish tray of the smoked fish machine**

The effort force needed is 0.51 N to lift the 5 kg of fish. The driven pulley has a 6 cm diameter with 11 teeth connected to it, and the other is 3 cm with 18 teeth connected to the stepper motor. The equation required to design the driven pulley The diameter and rpm of each pulley are computed using Equation 3.

$$\frac{d_1}{d_2} = \frac{n_2}{n_1} \tag{3}$$

Where,
d = diameter of the pulley and
n = angular velocity (radian per revolution) speed of the pulley.

The belt velocity (v) is computed using equation 4. The belt velocity (v) is computed using equation 4.

$$v = \frac{\pi d_1 n_1}{60} \tag{4}$$

The belt tension (F) is computed using equation 5.

$$F = \frac{P}{v} \tag{5}$$

Where P = transmitting power.

The belt length (L) is computed using equation 6.

$$L = \frac{\pi d_1}{2} + \frac{\pi d_2}{2} + 2D + \left(\frac{(d_1 - d_2)^2}{4}\right) \tag{6}$$

The torque (T) for the driven pulley and driver pulley is computed using equation 7.

$$\boldsymbol{T = \frac{P}{(\frac{2\pi n}{60})}} \tag{7}$$

The following are the calculated results for designing the mechanical mechanism of the machine. The pulley speeds of 6 cm and 3 cm are 25 rpm and 50 cm, respectively. The belt velocity is computed at 0.078 m/s, and the tension is 6.5 N. The belt length required is 34.4 cm, and the distance between the two pulleys is 10 cm at their center. The torque required for the driver and the driven pulley is 0.1 Nm and 0.2 NM, respectively. Therefore, a NEMA 23 stepper motor, which meets the required torque, is used.

### *2.3. Machine Description*

Figure 3 shows the machine's cross-sectional view of its major parts. This consists of the following:
1. Stepper motor
2. Food-grade stainless bowl
3. Boiling chamber part
4. Smoking chamber part
5. Electronic igniter
6. LPG tank

Figure 4 shows the top view, in which there are four exhaust fans at the boiling chamber and a smoke path going to the smoking chamber.

The stepper motor drives the gears and pulley installed on a vertically hanging 2.5x2x1-foot fish tray that can carry a fish batch with a maximum weight of 8 kg. With this setup, the machine can navigate between the boiling process and the smoking process, as shown in Figure 5.

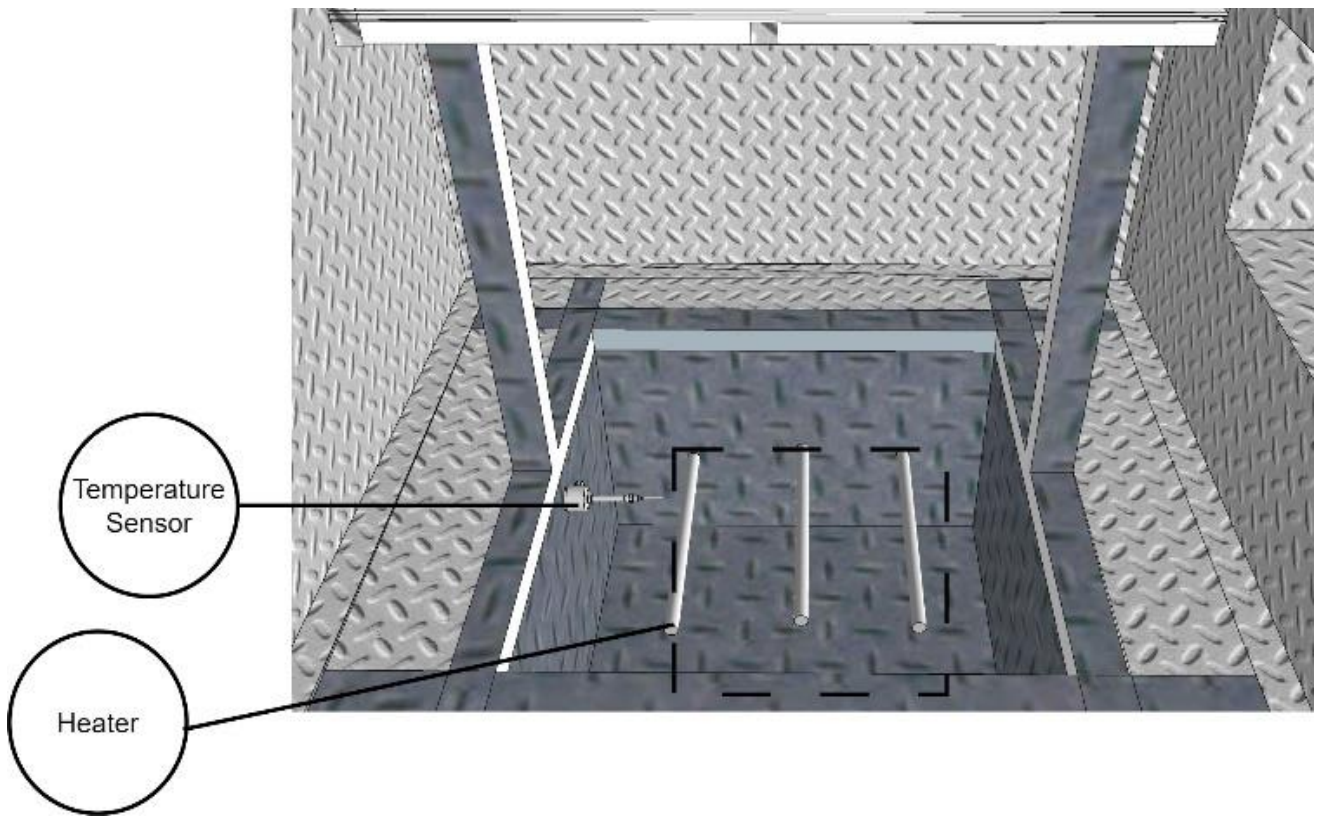

Fig. 6 Temperature sensor and heating elements

Figure 6 shows the components of the heating element and temperature sensor. Three heaters are installed to boil the water and cook the fish. The temperature sensor was a waterproof device submerged in water during the process and could withstand high temperatures. Figure 7 shows the controller box of the smoked fish machine. This consists of relays, a liquid crystal display (LCD), an electronic keypad, a pushbutton, a stepper motor driver, an Arduino Atmega328, a DC converter, and a power supply. For the smoke chamber, the components are the temperature sensor, exhaust fan, and igniter, as shown in Figure 8. All wires are fire-rated to withstand high temperatures.

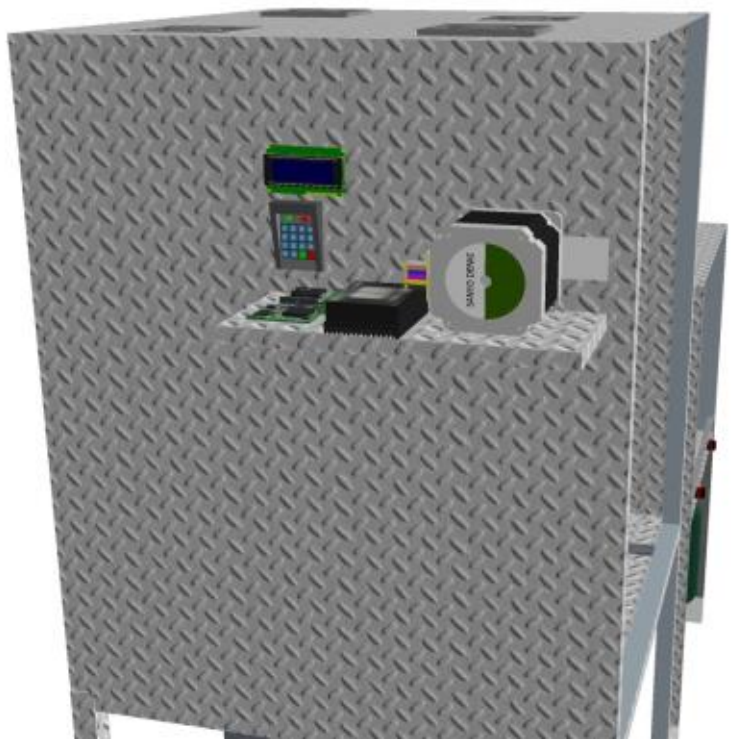

Fig. 7 Controller box of the smoked fish machine

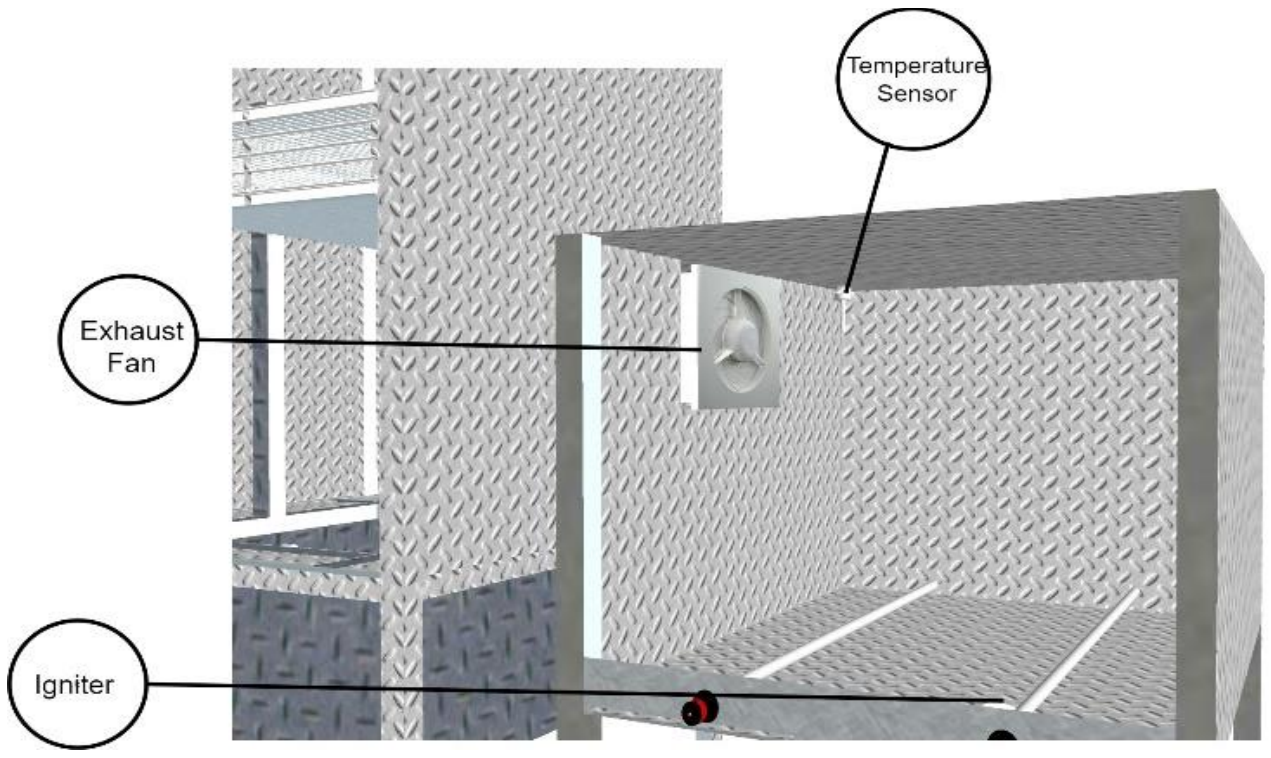

Fig. 8 Smoking Chamber consist of an Exhaust fan, Igniter and Temperature Sensor.

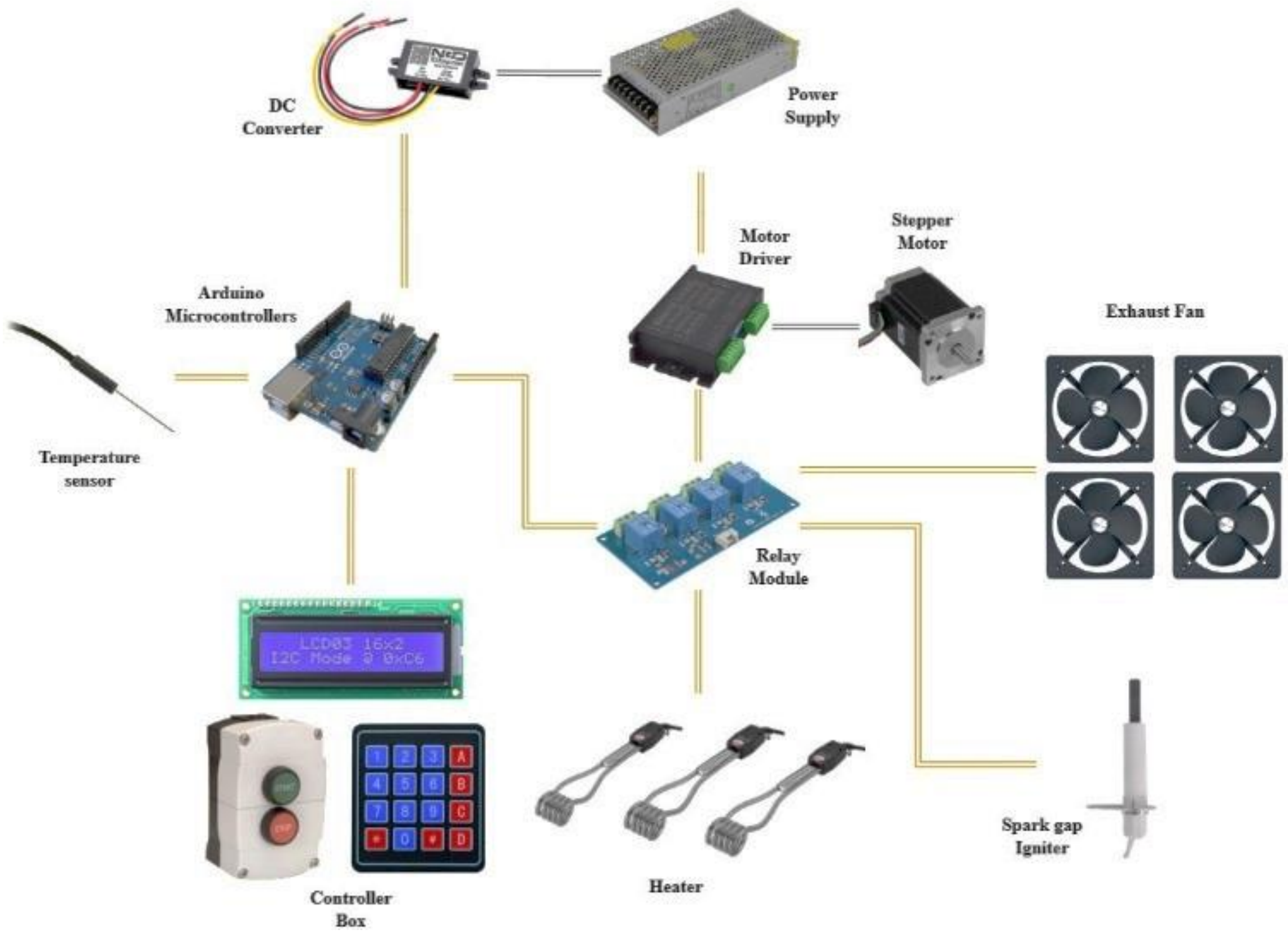

Fig. 9 System architecture of a smoked fish machine

## 3. Materials and Methods

The materials and methods section should contain details on the system architecture, principles of operation, and microcontroller programming of the proposed machine.

### *3.1. System Architecture*

In Figure 9, the machine consists of an Arduino microcontroller, heater, spark gap igniter, stepper motor with the motor driver, temperature sensor, exhaust fan, DC converter, relay module, power supply, and controller box.

### *3.2. Principle of Operation*

The microcontroller-based smoked fish machine controls the boiling temperature, smoke temperature, and operation time between boiling, smoking, and drying. At the start of the operation, the stepper motor will bring the fish tray down to the desired position, leveling the stainless bowl, and then the heater will start to energize to boil the water up to 100 °C. The temperature sensor regulates the desired temperature by turning on and off the heater control using a PID.

This is to maintain a temperature of 75 °C to 90 °C to cook the fish. The fish tray will return to its initial position after boiling time and start igniting the wood dust in the smoking chamber. Similar to the PID control in boiling temperature, the smoke temperature is regulated by the sensor, allowing the exhaust fan to turn on if it is above the desired level. This maintains a 60°C to 70°C temperature to smoke the fish. After smoking, the exhaust fan will turn on to dry the smoked fish until the operation is done.

### *3.3. Microcontroller Programming*

The microcontroller of the smoked fish machine used an Arduino ATMEGA328 to control the mechanism. The microcontroller is supplied by a 5V DC converter powered by a 12V switching power supply. The motor driver converts the electrical signal from the microcontroller to a mechanical signal received by the stepper motor.

The timer is built into the microcontroller system, which allows the user to set the time for the boiling, smoking, and drying of the smoked fish. PID is used to control the on-off of the heater and exhaust fan by reading the temperature of the DS18B20 temperature sensor.

This sensor is used to maintain the desired temperature according to user-set points. A 6-channel electronic relay module is used to have an on-off switch that is connected to a normally on contact. Figure 10 shows the flow chart of PID control for both the heater and exhaust fan.

Figure 11 shows the control program of the smoked fish machine. By using Matlab Simulink to program the PID control into the Arduino Mega hardware, they assign a general-purpose input/out (GPIO) pin for real-time monitoring of the temperature to display the response of the PID in the system during the boiling and smoking processes using scope.

The sensors, motor, and exhaust fans are linked to the Arduino interface. Setting the values of proportional (kp), integral (ki), and derivative (kd) could optimize the response of the system to control the temperature.

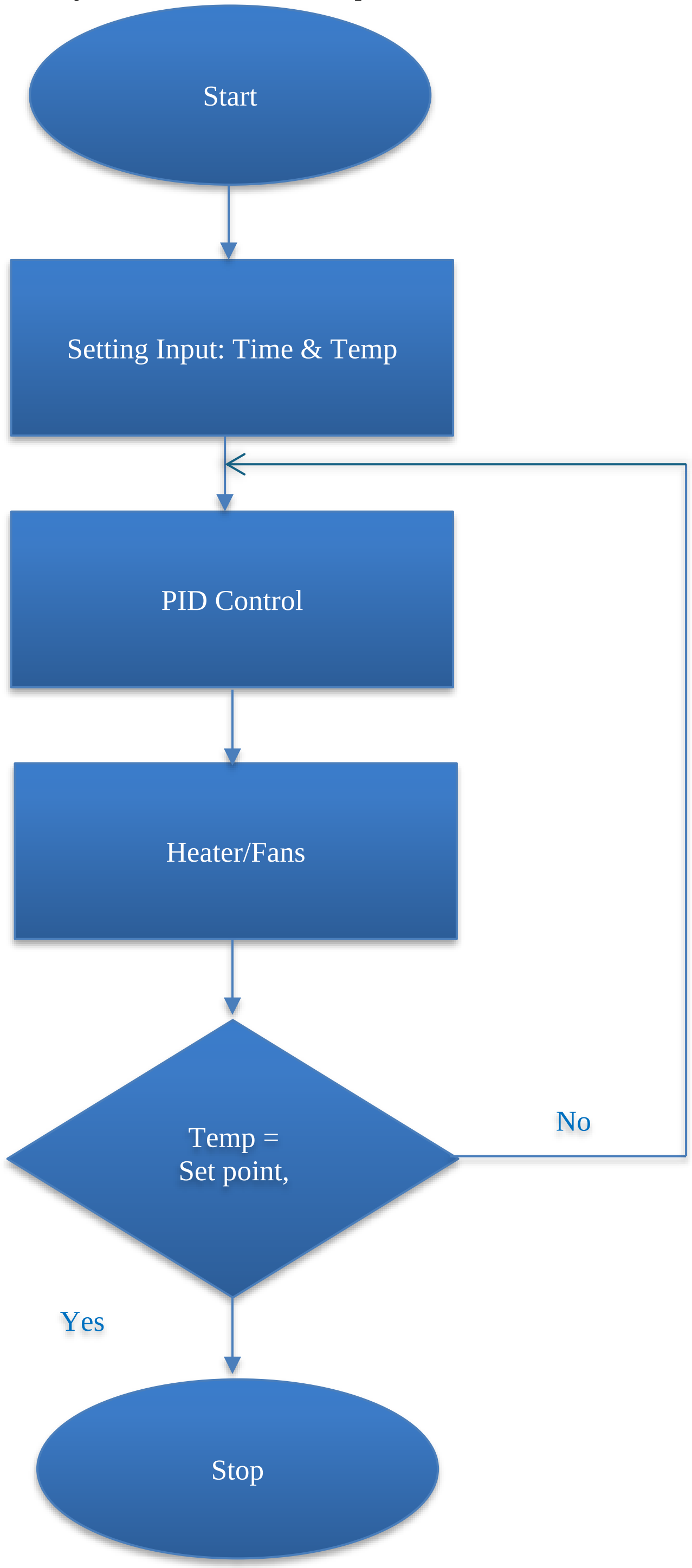

**Fig. 10 PID control of the smoked fish machine**

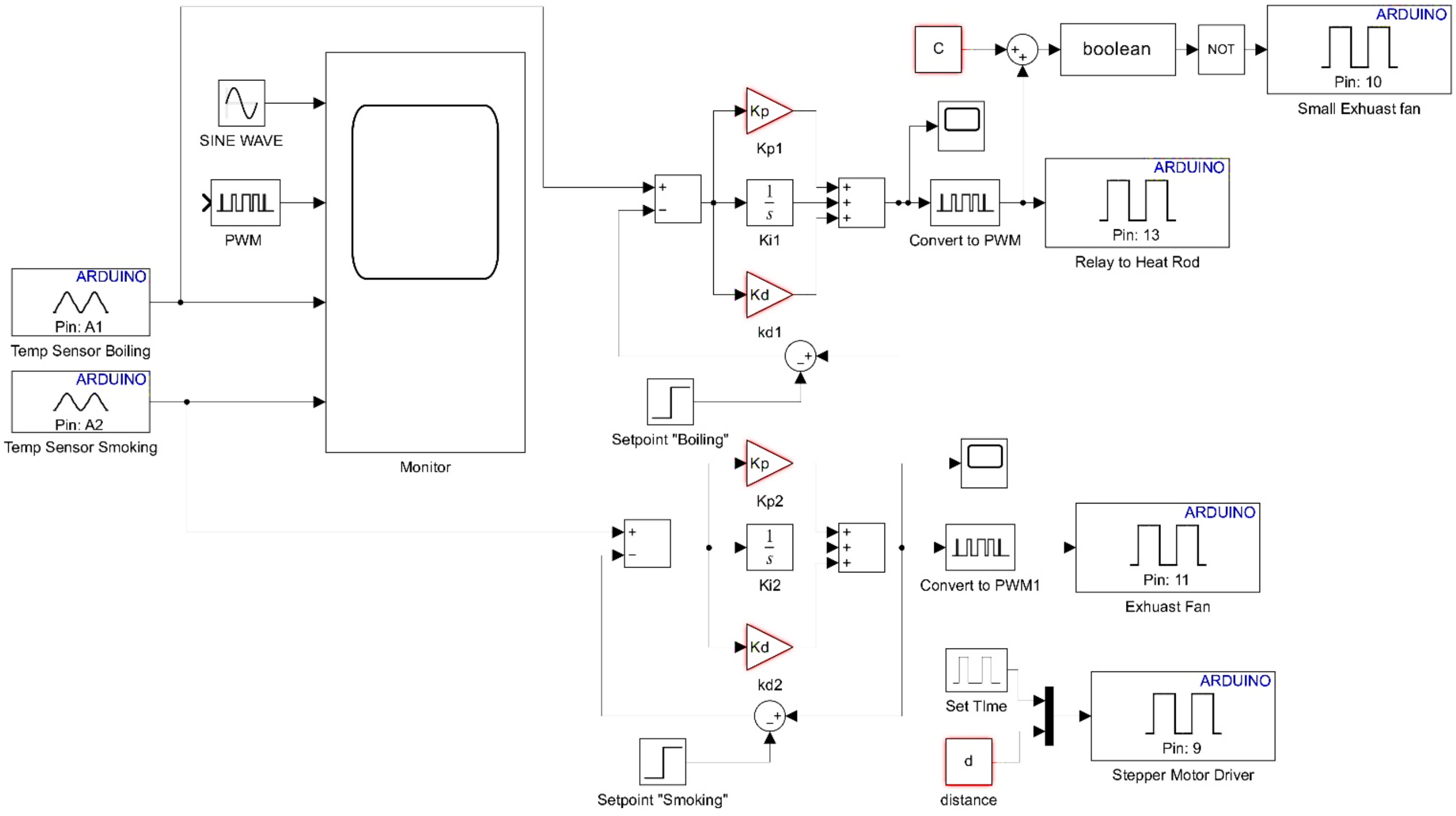

**Fig. 11 Control Program of the smoked fish machine**

## 4. Results and Discussion

### *4.1. Actual Machine Structure*

The microcontroller-based smoked fish machine was fabricated according to the design and mechanical specifications shown in Figure 12. The machine was able to operate and produce up to 5 kg of smoked fish in one whole process with the support of a driven pulley. In 60 minutes, the operation temperature can reach up to 98 °C for a process. Materials used in the machine can withstand high temperatures, and food-grade materials are used to minimize contamination and fish burning. Isolation is used for the protection of electrical wiring and electronic devices.

Table 1 shows the actual average setup time in each process for boiling water, cooking, smoking, and drying the smoked fish. The time of the process can vary based on the type and size of the fish. The production time efficiency increased from 4 hours using the traditional method to 1 hour using the proposed machine, which increased from 4 hours using the traditional method to 1 hour using the proposed machine.

Table 2 shows the testing of different types of fish by using the machine, setting the process by the amount of material used in the process with the same amount of fish weight, water, salt, and sawdust wood. The larger size of the scad has an average result of 61 minutes of operation. The scads are medium size, and the tilapia have the same result of 58 minutes of operation. The milkfish yielded a 65-minute result. This result shows the effect of the sizes and types of fish that vary the temperature and time of operation in the smoking process.

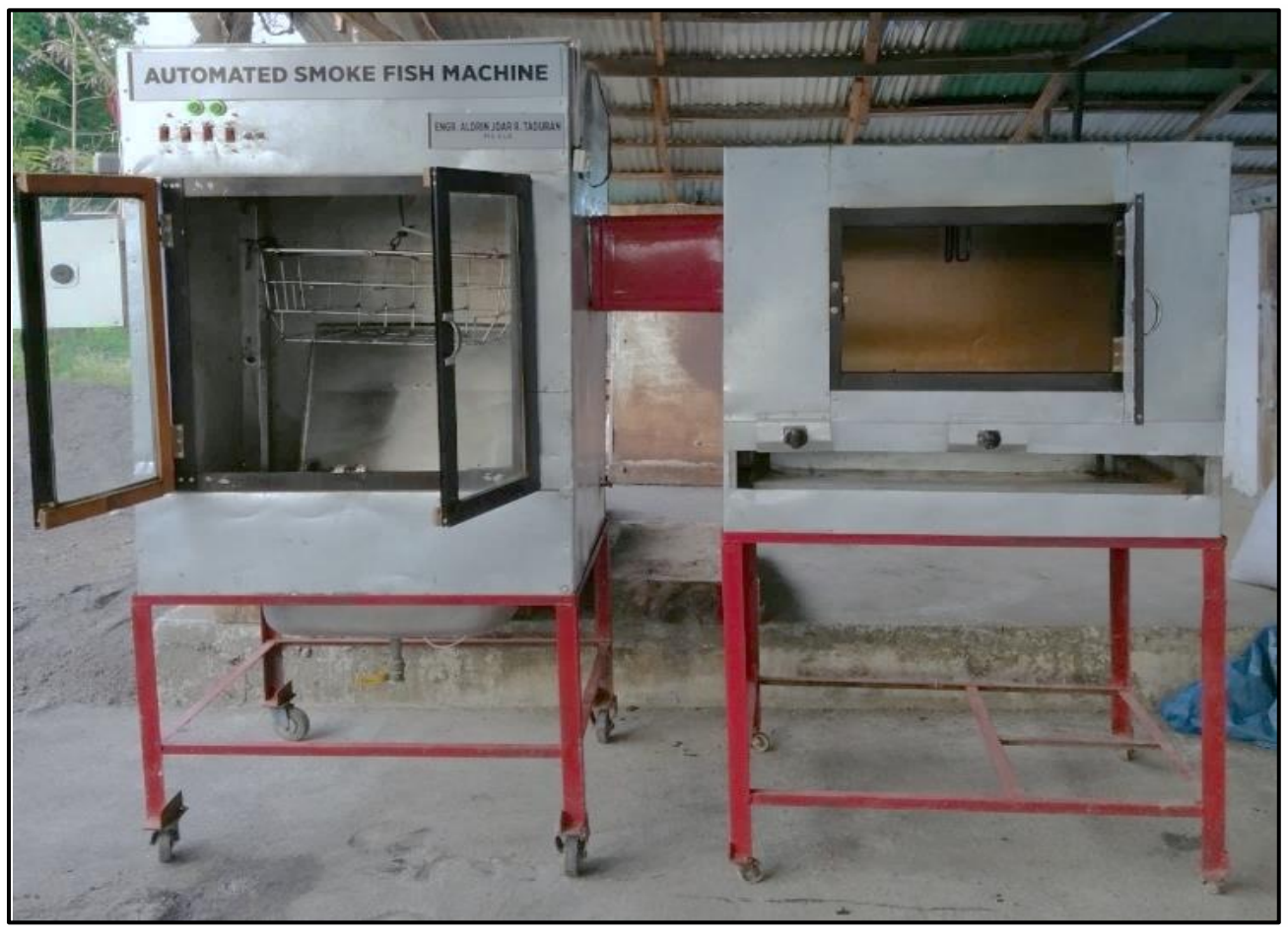

**Fig. 12 Actual microcontroller-based smoked fish machine**

**Table 1. Average time of the process of the smoked fish machine**

| Time of Water | Time of Cooking | Time of Smoking | Time of Drying |
|---|---|---|---|
| 5 minutes | 20 minutes | 15 minutes | 20 minutes |
| Total Time of Process | | 60 minutes/1hour | |

**Table 2. Average time of the process of the smoked fish machine**

| Type of Fish | Min-Max Temperature | Material | Time (minute) |
|---|---|---|---|
| Scads (Large) | Initial:29.76°C<br>Max:101.11°C | Water: 5gal<br>Salt: 6 kg<br>Sawdust: 2kg<br>Fish: 5kg | 61 |
| Scads (Medium) | Initial:28.17°C<br>Max: 98.65°C | Water: 5gal<br>Salt: 6 kg<br>Sawdust: 2kg | 58 |
| Milk Fish | Initial:29.76°C<br>Max: 99.01°C | Water: 5gal<br>Salt: 6 kg<br>Sawdust: 2kg<br>Fish: 5kg | 65 |
| Tilapia | Initial:28.92°C<br>Max:96.97°C | Water: 5gal<br>Salt: 6 kg<br>Sawdust: 2kg<br>Fish: 5kg | 58 |

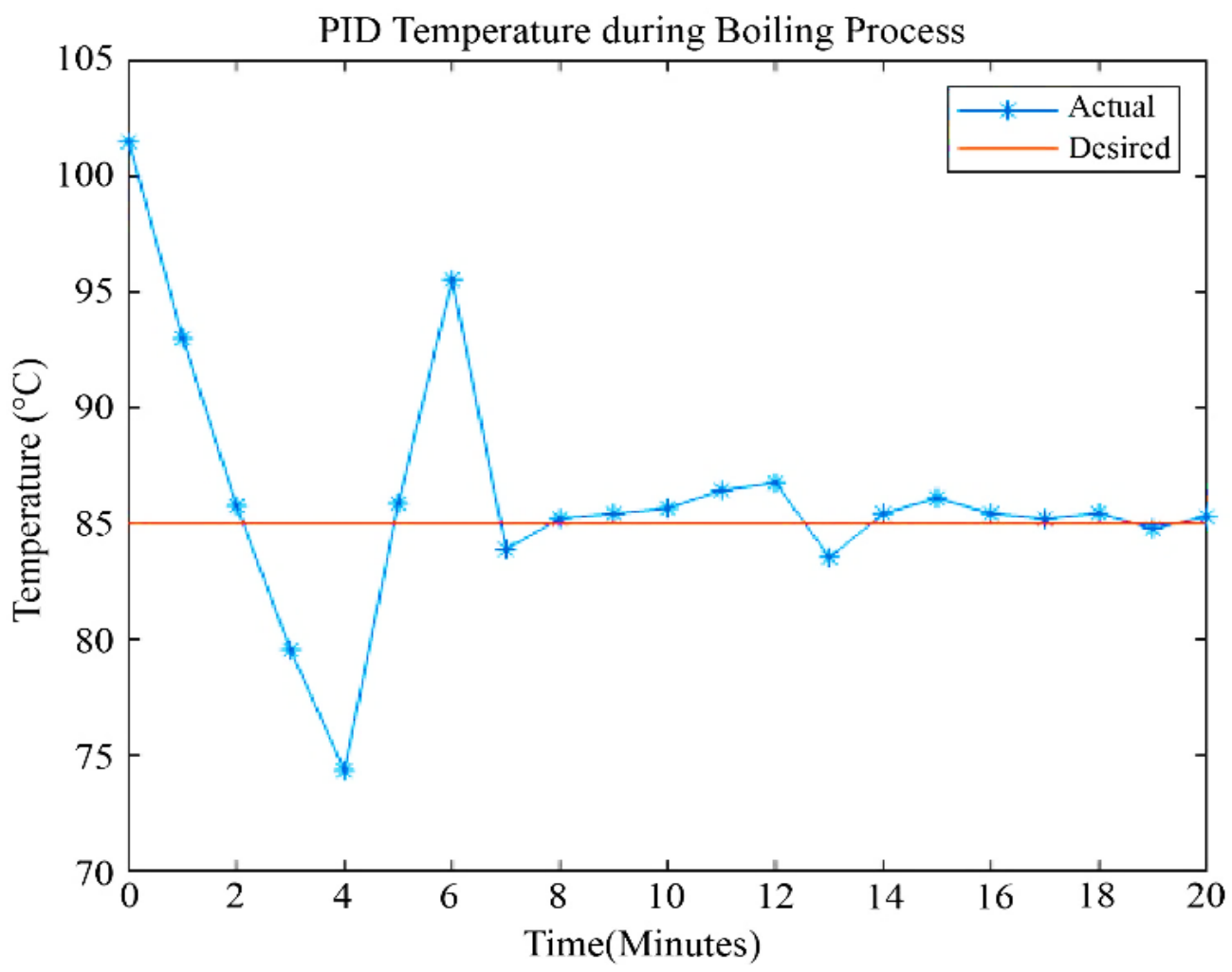

**Fig. 13 Graph of PID temperature during the boiling process**

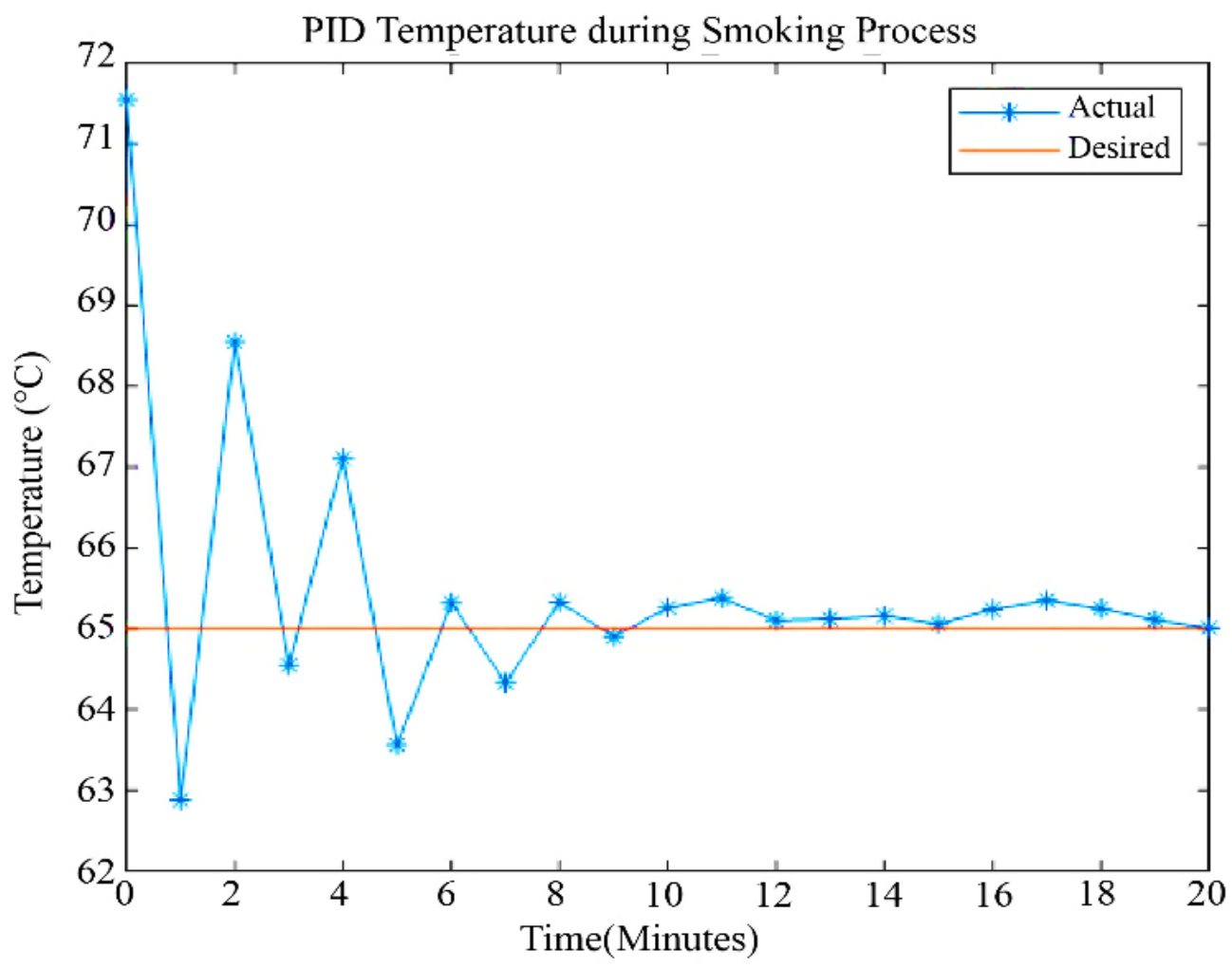

**Fig. 14 Graph of PID temperature during the smoking process**

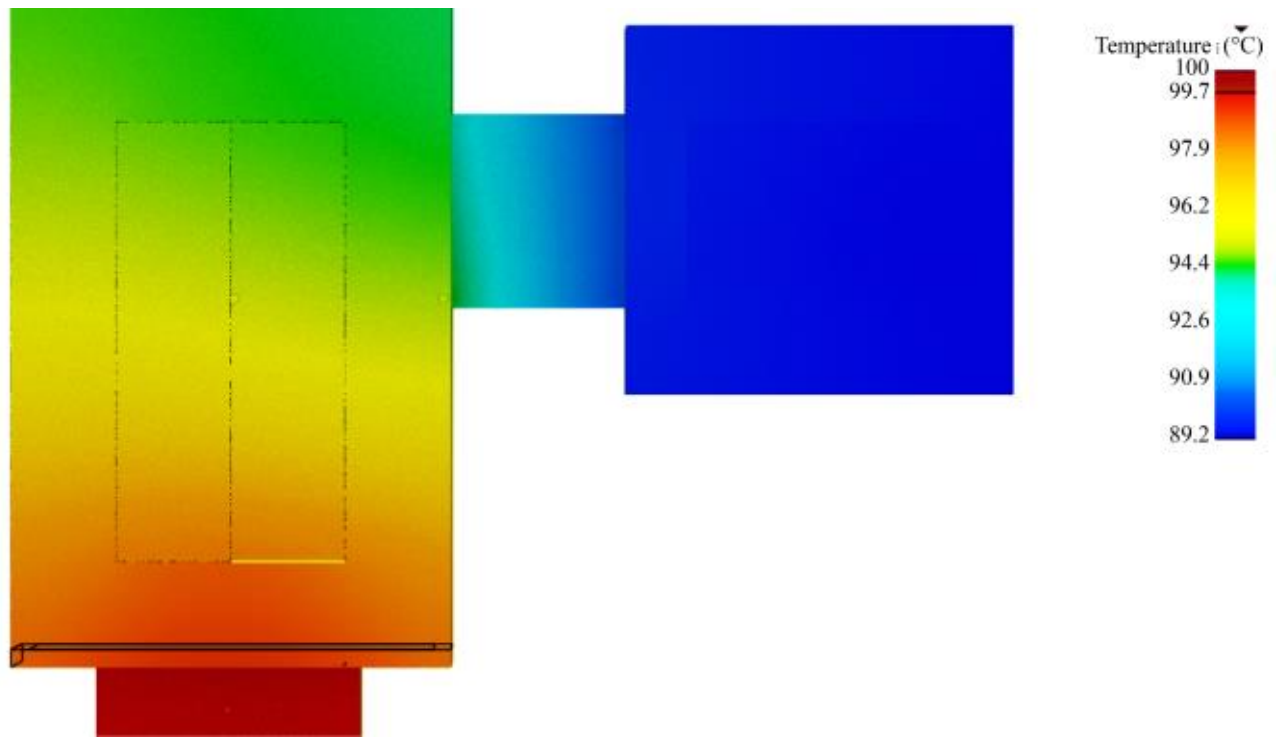

**Fig. 15 Heat Exchange during the boiling process**

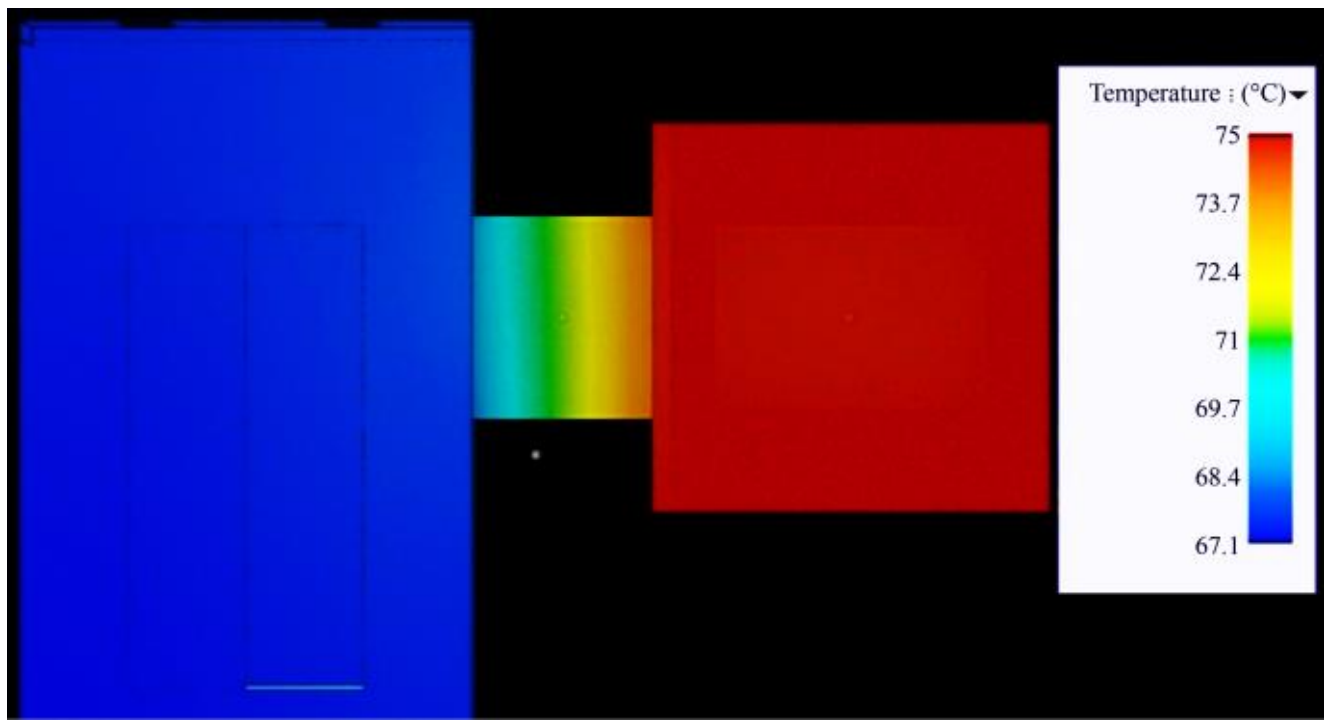

**Fig. 16 Heat Exchange during boiling smoking**

### 4.2. *Temperature During Boiling and Smoking*

Using PID controls as an optimizer for the machine's performance based on the desired temperature set point in the boiling and smoking processes. In Figure 13, during the 20-minute process of cooking the fish, the desired temperature setpoint is 85 °C. The PID controls the temperature from 85°C with a ±1°C read by the sensor. In Figure 14, during the 15-minute process of smoking the fish, the desired temperature setpoint is 65 °C. The PID controls the temperature from 65°C with a ±1°C read by the sensor. Using Ansys software to analyze and simulate the temperature distribution on the body of the smoked fish machine. In Figure 15, the heat exchange during the boiling process is shown by setting the temperature to a possible maximum temperature of 100 °C. The bottom surface reaches a temperature of 99.7 °C up to the upper surface, which reaches 94.4 °C. To reduce the heat dissipated, an outlet flow was needed with the help of an upper exhaust fan to reduce the machine's temperature.During the smoking process shown in Figure 16, the heat exchange on the smoking side reaches a temperature of 75 °C, while the smoke path has a temperature of 69.7 °C. The upper exhaust fan and exhaust inside the smoke chamber will help reduce the machine's temperature.

## 5. Conclusion

Developing a microcontroller-based smoked fish machine increased production efficiency from 4 hours to 1 hour with 5 kg of fish per batch over the traditional method.

The machine successfully handled the process of boiling, smoking, and drying fish simultaneously.

Based on the result, the use of PID in the machine for the temperature of the smoking and boiling processes has optimized the performance of controlling the temperature. This machine will be able to assist small and medium-sized smoked fish businesses in meeting consumer demand.

## Funding Statement

There is no funding grant for this research, and publication will be granted by the university after it is accepted in a journal.

## Acknowledgments

Without the university's unwavering support and encouragement, this research project will not be finished.

## Appendix 1

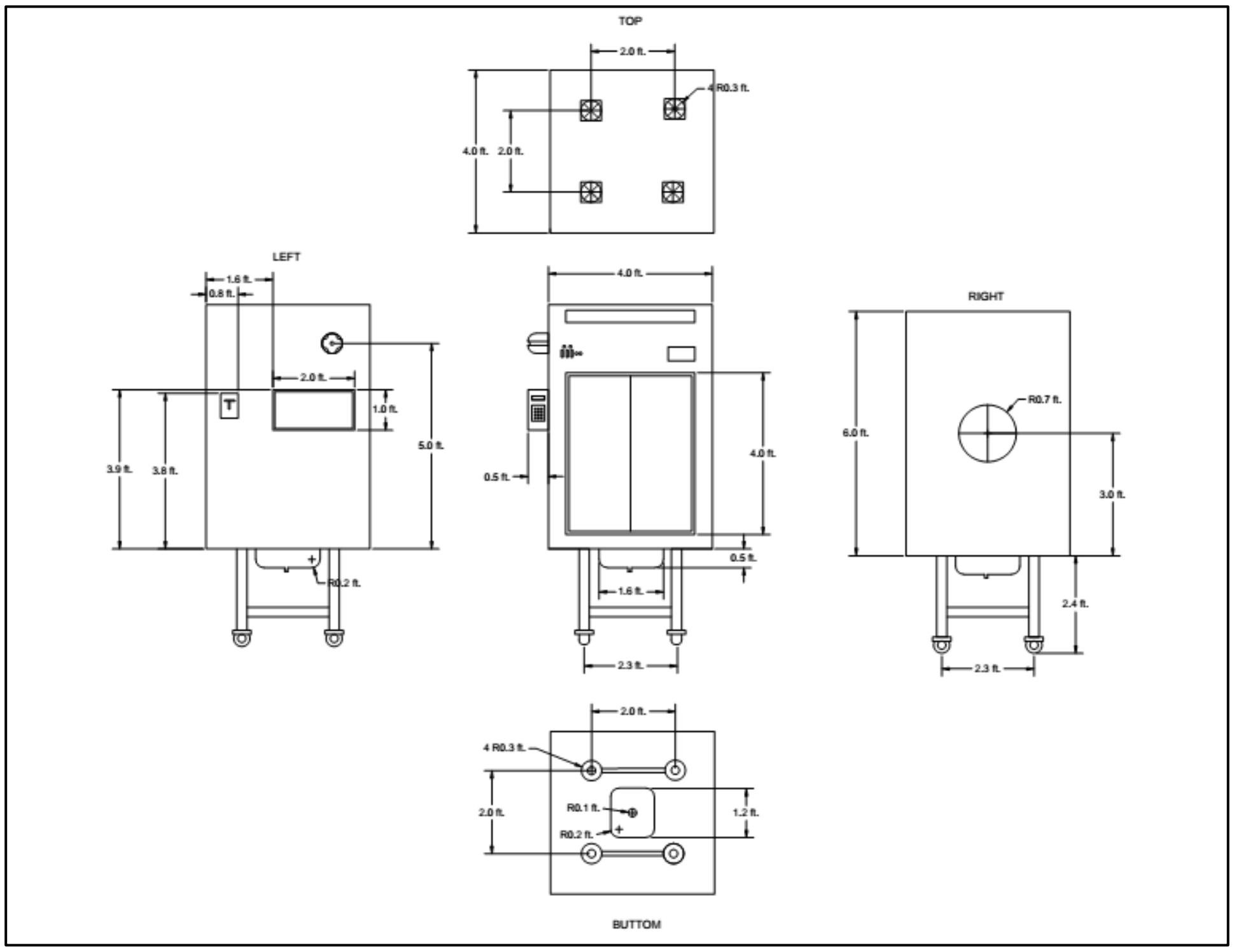

## Appendix 2

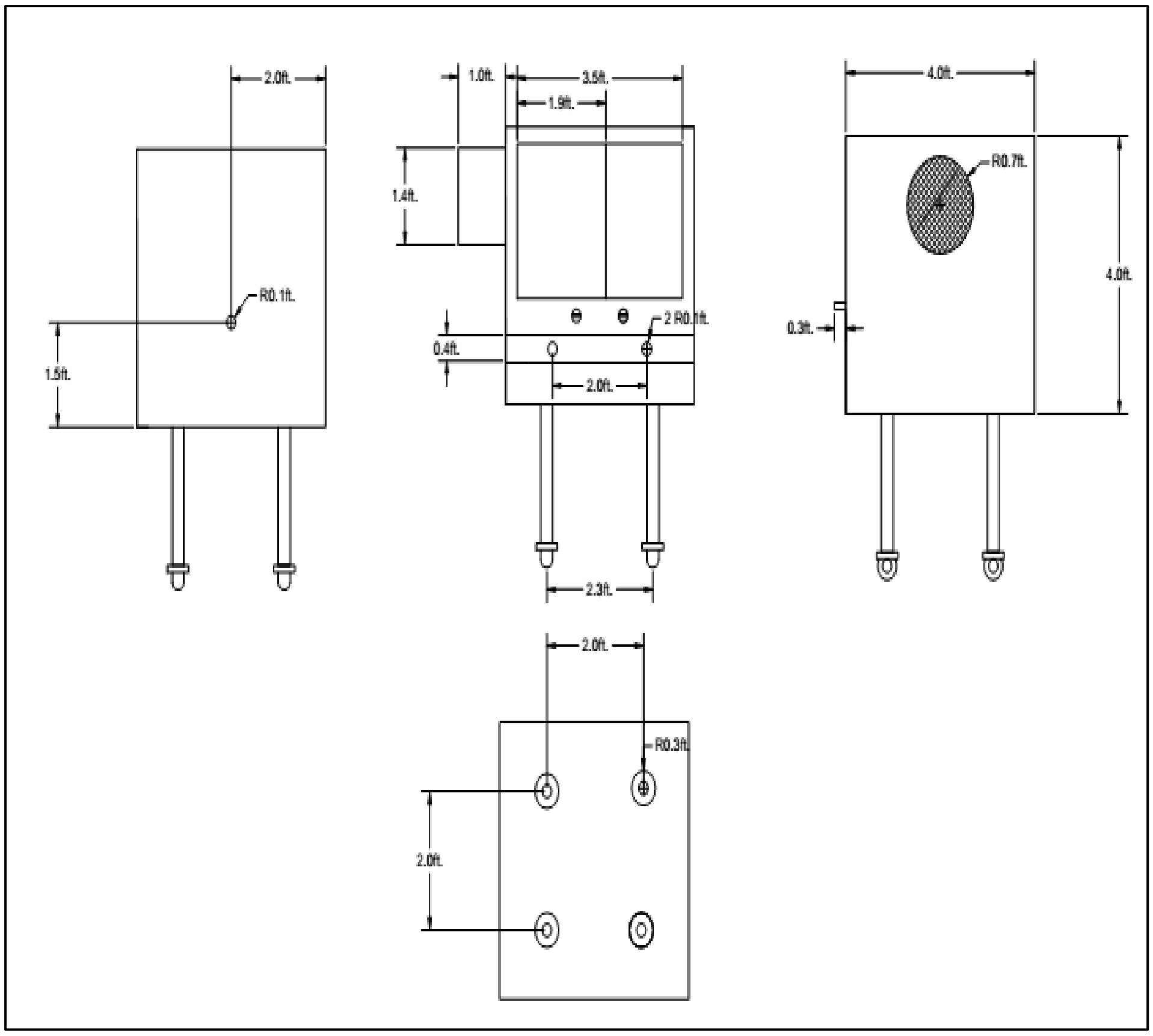

## Appendix 3

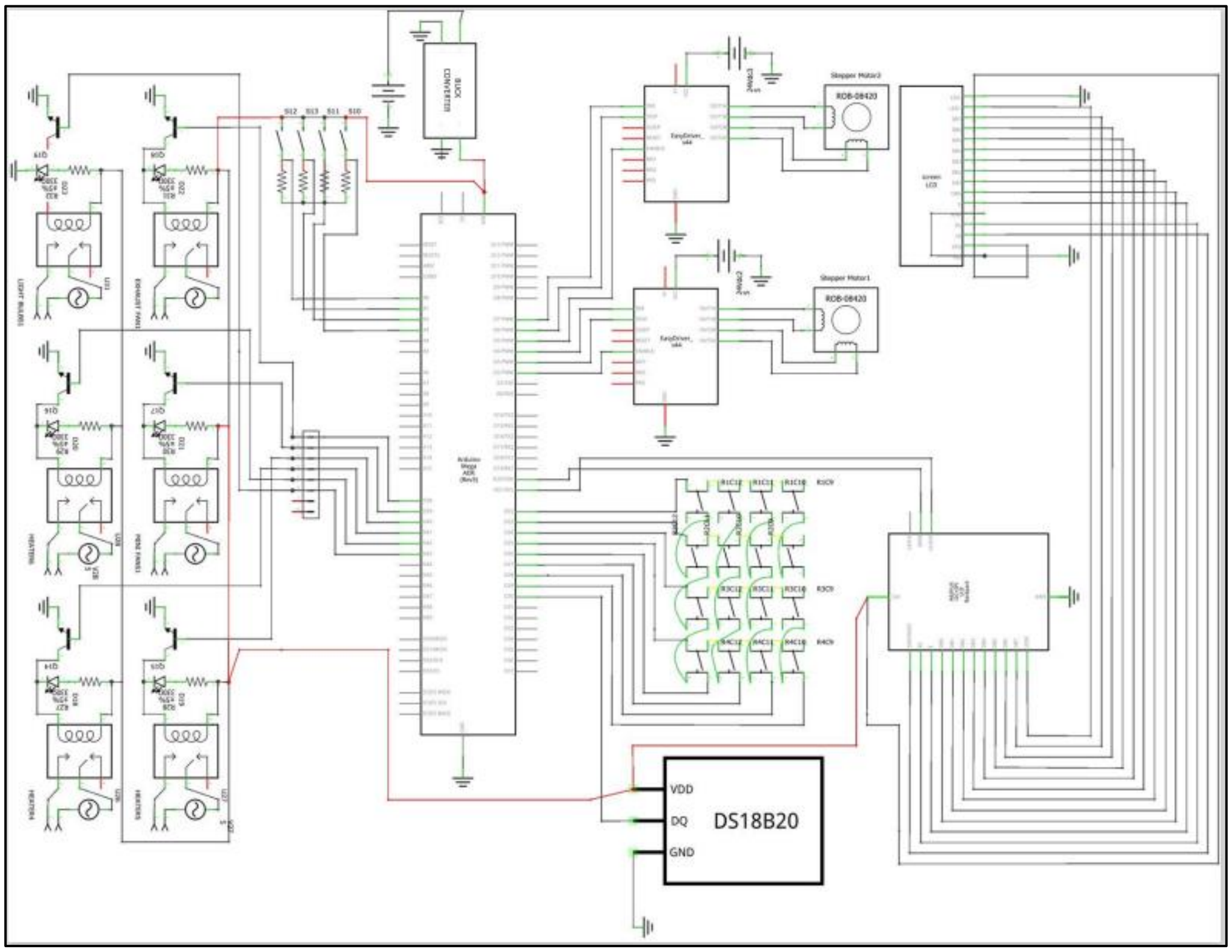

## Appendix 4

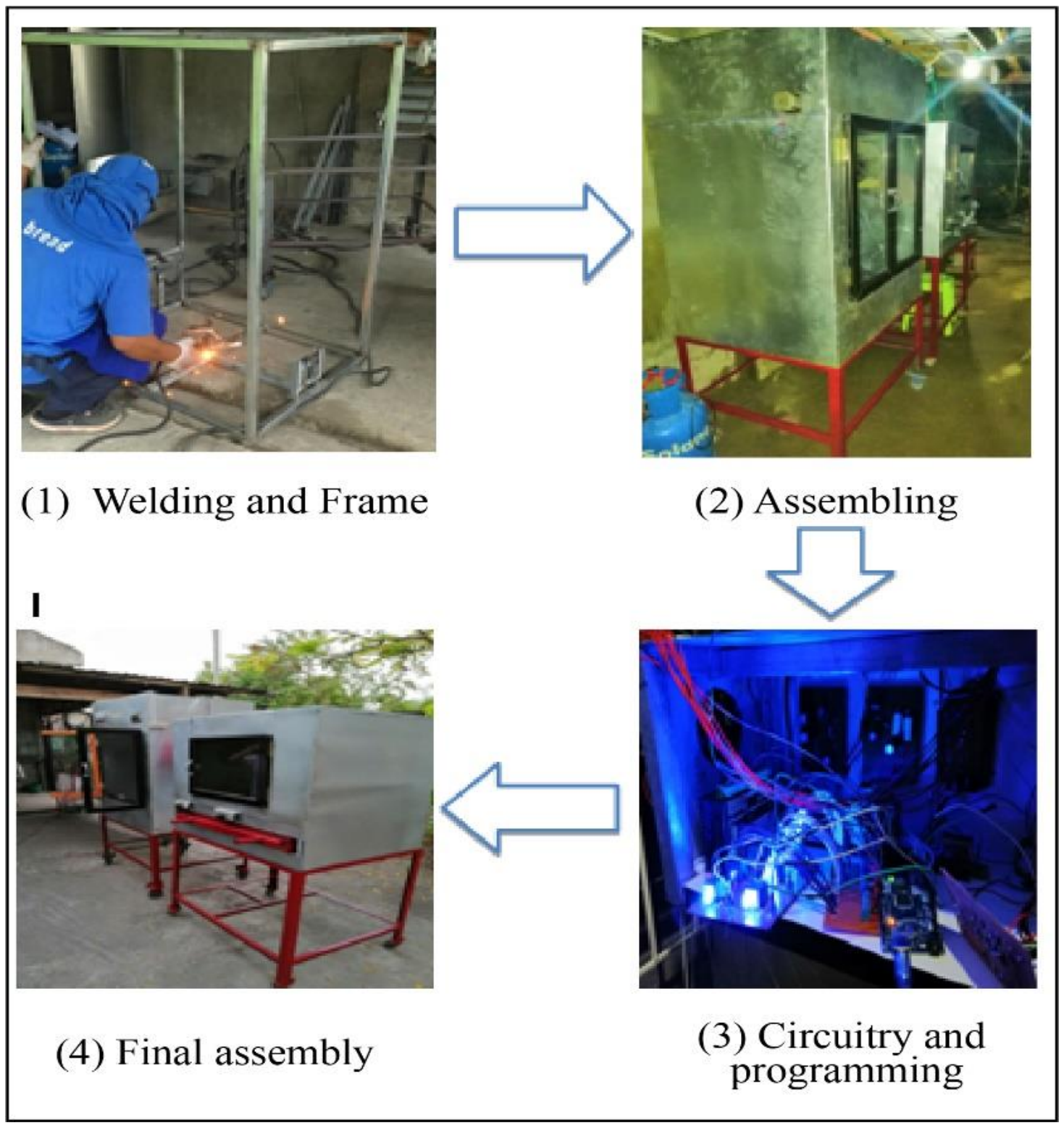